\documentclass[aps,prb,amssymb]{revtex4}

\usepackage{blindtext} 
\usepackage{graphicx}
\usepackage{bm}

\def\Lm {\Lambda}

\def\de {\delta}
\def\si {\sigma}
\def\ep {\epsilon}

\def\ef {\epsilon_F}
\def\be {\begin{equation}}
\def\ee {\end{equation}}
\def\bea {\begin{eqnarray}}
\def\eea {\end{eqnarray}}

\def\om {\omega}

\begin{document}


\centerline{Published in Journal of Low Temperature Physics,
  {\bf 191}, 123 (2018), https://doi.org/10.1007/s10909-017-1847-2}

\title{Conductivity of weakly disordered
metals close to a ``ferromagnetic'' quantum critical point }

\author{George Kastrinakis }

\affiliation{                    
   Institute of Electronic Structure and Laser (IESL), 
Foundation for Research and Technology - Hellas (FORTH), 
P.O. Box 1527, Iraklio, Crete 71110, Greece$^*$ }

\date{24 November 2017}

\begin{abstract} 
We calculate analytically the conductivity of weakly disordered metals close
to a ``ferromagnetic'' quantum critical point in the low temperature regime.
Ferromagnetic in the sense that the effective carrier potential $V(q,\om)$,
due to critical fluctuations, is peaked at zero momentum $q=0$.
Vertex corrections, due to both critical fluctuations and impurity scattering, 
are explicitly considered. We find that only the vertex corrections 
due to impurity scattering, combined with the self-energy, generate
appreciable effects as a function of the temperature $T$ and the control 
parameter $a$, which measures the proximity to the critical point.
Our results are consistent with resistivity experiments in several materials
displaying typical Fermi liquid behavior, but with a diverging prefactor
of the $T^2$ term for small $a$. 

\end{abstract}

\maketitle
 
\vspace{0.6cm} 

{\bf 1. Introduction}

\vspace{0.3cm} 

Itinerant electron systems display non-trivial behaviour close to 
a quantum critical point (QCP). E.g. some observables may diverge
upon approaching the QCP. 
Our work is motivated by a number of experiments on several materials 
\cite{pag,bia,gri,geg1,geg,but,shib,bali,naka,anal}, which display 
typical Fermi liquid (FL) behaviour for appropriately low temperature
$T$. That is,
{\em quadratic} in $T$ resistivity and {\em linear} in $T$ specific heat.
These materials include CeCoIn$_5$ \cite{pag,bia},
Sr$_3$Ru$_2$O$_7$ \cite{gri}, YbRh$_2$Si$_2$ \cite{geg1,geg},
La$_{2-x}$Ce$_x$CuO$_4$ \cite{but}, Tl$_2$Ba$_2$CuO$_{6+x}$  \cite{shib},
CeAuSb$_2$ \cite{bali}, YbAlB$_4$ \cite{naka}
and BaFe$_2$(As$_{1-x}$P$_x$)$_2$ \cite{anal}.
However, the prefactors of these quantities {\em diverge} in the
vicinity of the respective QCP's as power laws of the criticality parameter
$a$, which measures the proximity to the QCP. $a$ may be determined
by the electron filling factor, the pressure,
or the magnetic field $H$ (which is related to filling, through the Zeeman
term) \cite{revi,gkfluc}. The $T^2$ resistivity appears within various
material and $H$ dependent ranges. E.g. up to 70 mK for
YbRh$_2$Si$_2$ \cite{geg1}, between 0 - 1.2 K for $H$=5-14 T, respectively,
for CeCoIn$_5$ \cite{pag},
for up to 10 K and $H \leq 1$ T for Sr$_3$Ru$_2$O$_7$
\cite{gri}, up to 15 K at $H=25$ T for CeAuSb$_2$ \cite{bali}, and
up to 100 K at $H=45$ T for  Tl$_2$Ba$_2$CuO$_{6+x}$  \cite{shib}.
It is possible that in this regime $T$ is less or at most of the order
of the impurity scattering rate $\tau_o^{-1}$.

We have shown in \cite{gkfluc}, via analytic diagrammatic calculations,
that this critical FL behaviour can be consistently understood as arising
from the exchange of relevant ferromagnetic fluctuations with {\em small}
momentum $q$ among the quasi-particles.
Our approach assumes that we deal with weakly disordered metallic
systems. Herein we extend our previous calculation of the conductivity in the 
low $T$ regime, via a more comprehensive inclusion of vertex corrections. 
The latter are due both to the fluctuation potential $V(q,\om)$ and to elastic
(spinless) disorder scattering. The part of vertex corrections due to 
$V(q,\om)$ yields no essential modifications on the results 
already obtained in \cite{gkfluc}.

\vspace{0.6cm} 

{\bf 2. The model}

\vspace{0.3cm} 

Henceforth, all momenta are 3-D or 2-D vectors, 
though we do not use bold letters. We consider the Green's function 
\be
G_o^{R,A}(k,\ep)=\frac{1}{\ep-\xi_k \pm i /2\tau_o} \;\; , \;\;
\xi_k = \ep_k -\ef  \;\; , \;\;
\ee
with $\ep_k$ the quasiparticle dispersion, $\ep_F$ the Fermi energy,
and $\tau_o$ the momentum relaxation time due to impurities. In the 
weak disorder regime \cite{agd,lee} $\ep_F \tau_o \gg 1$.
$\tau_o^{-1}$ is {\em important as a regulator} in the calculations.
In fact, the characteristic FL $T^2,\ep^2$ dependence of Im $\Sigma$ in 
eq. (\ref{scat}) is due to the finite $\tau_o^{-1}$.

The dominant electron-electron interaction is assumed to be
the ``ferromagnetic'' fluctuation potential (or fluctuation propagator)
\cite{gkfluc,her,mil1} peaked at $q=0$
\be
V(q,\om)=\frac{g}{-i \om /(D q^2+r)+  \xi^2 q^2 + a} \;\;,
\ee
with $g$ the coupling constant, $\xi$ the 
correlation length and $a$ measuring the distance from 
the QCP. The criticality parameter $a$ depends 
on e.g. $H$, as in the systems of interest mentioned 
below, like $a=h^s$, $h=|H/H_c-1|$, $s>0$, where $H_c$ is the critical field.
The factor $D q^2$ indicates disorder induced diffusion of the 
quasiparticles, with diffusion coefficient $D$ \cite{lee,gkdef}.

For the purpose of our calculations, we will 
treat $\xi$ and $a$ as {\em independent parameters}. This procedure, also
followed in \cite{gkfluc}, is entirely consistent, as can be 
seen from the details of the calculations below. Also, after eq. (\ref{r1ga}),
we discuss the role of
the Gaussian regime $\xi^{2} \; a= \text{const.}$ \cite{revi,her,mil1}, 

We have shown in \cite{gkfluc} that, for the 
self-energy $\Sigma = \text{Tr }G_o \; V$,
the quasi-particle scattering rate is 
\bea
\text{Im }\Sigma(x,a) = F_d(a,\xi) \; x^2 \;\; , \;\; 
x=\text{max}\{T,\ep\}  \;\; .\;\;  \label{scat}
\eea
Here $F_d(a,\xi)$ scales like a {\em negative}
power of the criticality parameter $a$ in $d=2,3$ dimensions.
We obtained $F_d \propto a^{-2} \;
\left[ \ln \left( \xi q_{max}/\sqrt{a} \right) -1\right]$ for $r=0,D>0$,
and $F_d \propto a^{-1} \; \xi^{-2}$ for $r>0,D=0$. This result can be also
considered in the frame of the Gaussian regime, though it was derived
without assuming any dependence between $\xi$ and $a$.
In Appendix A we explicitly derive the result corresponding
to eq. (\ref{scat}) for the case $\ep > T$.

In the following, we consider the {\em total} quasi-particle scattering rate
\be
2 S \equiv \tau^{-1}(T,a) = \tau_{o,i}^{-1} + 2\; \mbox{Im} \; \Sigma(\ep=0,T,a) 
\;\;, \;\;   \label{sigm}
\ee
with $\tau_{o,i}^{-1}$ due to impurity scattering. 
Then the Green's function is taken as 
\be
G^{R,A}(k,\ep)=\frac{1}{\ep-\xi_k  \pm i S} \;\; , \;\;  \label{grs}
\ee
i.e. it includes the self-energy of eq. (\ref{sigm})
due to the fluctuation potential $V(q,\om)$. 

\vspace{0.6cm} 

{\bf 3. Calculation of the vertex corrections}

\vspace{0.3cm} 

We wish to calculate the conductivity $\si$, by including vertex
corrections. Our treatment is similar to the one of Mahan \cite{mahan}
for electron-phonon scattering. However, ours is different in a number of
aspects, due to the different $V(q,\om)$ and $G(k,\ep)$ considered here,
the scattering by impurities,
the specific functions $f(\ep),n(\om)$ defined below etc.
Dell'Anna and Metzner \cite{metz} have treated the conductivity with 
vertex corrections for a scattering potential similar to our $V(q,\om)$.
However, disorder is {\em not} included in their Green's function, our
self-energy differs from theirs (while
d-wave form factors are included in their potential), and our
results differ significantly (this is also due to the different 
approximations made).
$\si$ is given by \cite{mahan}
\bea
\si = \frac{2 \;e^2}{3} \; \lim_{\om_0 \rightarrow 0} 
\frac{\text{Im} \; \Pi(\om_0)}{\om_0}
\;\;, \;\; 
\eea
where we analytically continue $i \om_l \rightarrow \om_0$ in
\bea
\Pi(i \om_l) = T \sum_{\ep_n} \sum_k \;  {\bf v}_k
\; {\bf \Gamma}(k, i\ep_n, i\ep_n+i\om_l) \; G(k,i\ep_n+i\om_l) \; G(k,i\ep_n)  
\;\; . \;\;
\eea
C.f. fig. 1.
Here $e$ is the charge of the electron and $ {\bf v}_k = {\bf \nabla}_k \ep_k$. 
The {\em vector} vertex function 
${\bf \Gamma} (k,i \ep_n,i \ep_n+i \om_l)$ (with $\om_l$
the energy difference between upper and lower lines) depends on the 
interactions - c.f. below. 
We consider scattering {\em both} via $V(q,\om)$ and from the impurities.
Here the Matsubara energies are $\ep_n=(2n+1)\pi T, \om_m=2\pi m T$ and
$ \om_l=2\pi l T$.
 
\begin{figure}[tb]
  \includegraphics[width=4truecm]{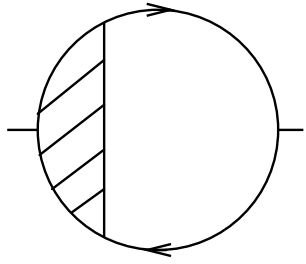}
\caption{Feynman diagram for the conductivity. 
The continuous
lines are the fermion propagators, i.e. the Green's function of 
eq. (\ref{grs}). The vertex function ${\bf \Gamma}$ is on the left of the
bubble.}

\label{fig.1}
\end{figure}

With $f(\ep)=(1/2)\tanh(\ep/2T)$ \cite{fnw} and $\de \rightarrow 0^+$, 
it can be shown that
\bea
\si= \frac{e^2}{3 \; \pi} \;
\int_{-\infty}^{+\infty} \; d\ep \; \frac{df(\ep)}{d\ep} \; \sum_k \; {\bf v}_k
\left \{
{\bf \Gamma}(k, \ep-i\de, \ep+i\de) \; G^R(k,\ep) \; G^A(k,\ep)  
- \text{Re}\left[ {\bf \Gamma}(k, \ep+i\de, \ep+i\de) \; 
\left(G^R(k,\ep)\right)^2
\right ] \right \}
\;\; . \;\;
\label{condsi}
\eea
This expression contains {\em two different variants} of the vertex
function, with different energy arguments. Writing
\be
{\bf \Gamma}(k, \ep+i\de, \ep+i\de) = {\bf v}_k \; \Lm(k, \ep+i\de, \ep+i\de)
\;\; ,\;\;  
\ee
and using the Ward relation $\Lm(k, \ep+i\de, \ep+i\de)=
1+(\partial/\partial \xi_k) \Sigma(k,\ep)$ (c.f. ref. \cite{mahan}, 
eq. (7.1.27) and after eq. (7.3.4)), we obtain
\be
\Lm(k, \ep+i\de, \ep+i\de) = \Lm(k, \ep-i\de, \ep-i\de) = 1 \;\;.\;\;
\label{lapp}
\ee
As mentioned in \cite{gkfluc}, after eq. (14), the dependence of
Im $\Sigma$ on $k$ is negligible for $k$ within a thick layer around the Fermi
momentum $k_F$.
We note that ${\bf \Gamma}(k,\ep-i\de, \ep+i\de)$ is {\em not} given by a Ward 
identity \cite{mahan}. To calculate it, we turn to the respective ladder 
diagram approximation, {\em without} crossing interaction lines, in which
${\bf \Gamma}(k,i \ep_n,i \ep_n+i \om_l)$ obeys the equation shown in fig. 2
\bea
{\bf \Gamma}(k,i \ep_n,i \ep_n+i \om_l) = 
{\bf \Gamma}^0(k,i \ep_n,i \ep_n+i \om_l)
+ n_i \sum_q U_i(q)^2 \; G(k+q,i \ep_n +i \om_l) \; 
G(k+q,i \ep_n) \; {\bf \Gamma}(k+q,i\ep_n,i\ep_n+i \om_l)  \nonumber \\
+T \sum_q \sum_{\om_m} V(q,i \om_m) \; G(k+q,i \ep_n+i \om_m+i \om_l)\; 
G(k+q,i \ep_n+i\om_m ) \; {\bf \Gamma}(k+q,i\ep_n+i\om_m, i\ep_n+i\om_m+i \om_l) 
\;\; . \;\;   \label{lamb}
\eea
$U_i(q)$ is the impurity scattering potential and $n_i$ the concentration
of impurities.

The relevant Aslamazov-Larkin (AL) diagrammatic contribution to the vertex
${\bf \Gamma}$ has been discussed in refs.
\cite{chu1,slee}. However, it was shown that for the charge vertex, and 
in the $q=0$ limit, where $q$ is the momentum difference of the two fermion
lines at the vertex, the AL contribution vanishes. Hence we do not consider 
it here.

\begin{figure}[tb]
  \includegraphics[width=10truecm]{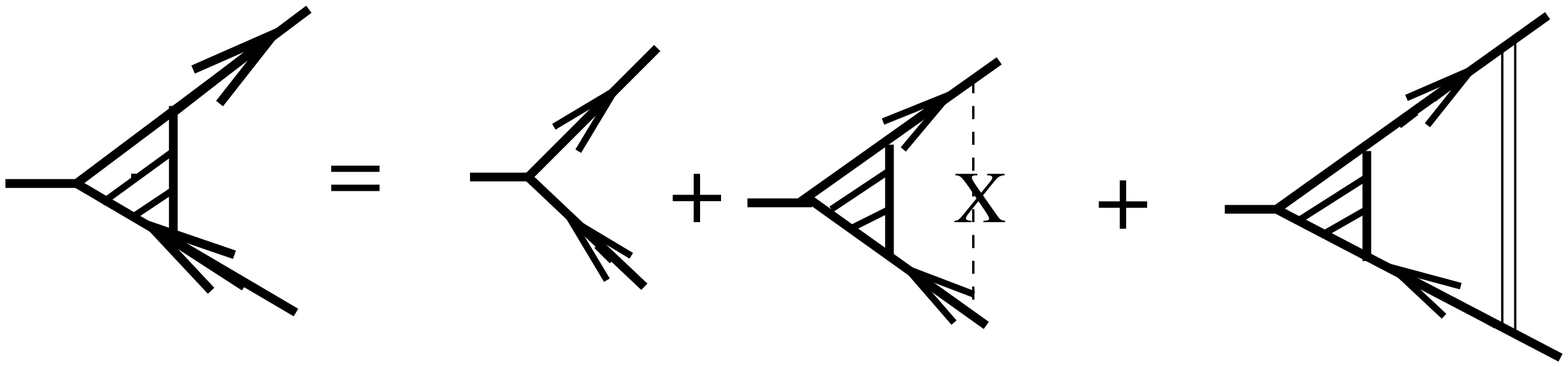}
\caption{Ladder diagrams for the vector vertex function ${\bf \Gamma}$. 
The dashed line with a cross stands
for impurity scattering and the double line on the right for the potential
$V(q,\om)$. }

\label{fig. 2}
\end{figure}

We make the usual assumption that 
\bea
{\bf \Gamma}(k,i \ep_n,i \ep_n+i \om_l)= {\bf v}_k \; 
\Lm(k,i \ep_n,i \ep_n+i \om_l)  
\;\; ,\;\;
{\bf \Gamma}^0(k,i \ep_n,i \ep_n+i \om_l) = {\bf v}_k \;\; , \;\; 
\eea
i.e. the vector dependence is just given by $ {\bf v}_k $.

For the solution of eq. (\ref{lamb}), we first look at the term 
involving $V(q,\om)$
\bea
W = T  \sum_{\om_m} V(q,i \om_m) \; G(k+q,i \ep_n+i \om_m+i \om_l)\; 
G(k+q,i \ep_n+i\om_m ) \; \Lm(k+q,i\ep_n+i\om_m, i\ep_n+i\om_m+i \om_l) 
\;\; . \;\;
\eea
In order to evaluate it, we consider $n(\om)=(1/2)\coth(\om/2T)$ \cite{fnw}, 
and the function of the complex variable $z$
\bea
 F(z) = n(z) \; V(q,z) \; G(k+q,i \ep_n+ z +i \om_l)\; 
G(k+q,i \ep_n+z ) 
\; \Lm(k+q,i\ep_n+z, i\ep_n+z+i \om_l)   \;\;. \;\;
\eea
Then we apply Cauchy's residue theorem for a closed contour $C$ 
at infinity, thus obtaining
\bea
\frac{1}{2 \pi i}  \oint_C dz \; F(z) = W+I_V+I_1+I_2+R_V+R_1+R_2+L= 0 \;\;.\;\;
\label{eqint}
\eea
The integrals $I_V,I_1,I_2$ are along the branch cuts of $V$ and the two $G$'s,
and are given below.
$R_V,R_1,R_2$ are the residues of $F(z)$ due to the poles of 
$V$ and the two $G$'s respectively. They are negligible, as discussed in
Appendix B. $L$ is the contribution from the poles of $\Lm$, which will
also turn out to be negligible, as shown in Appendix B. 

We have
\bea 
I_V = \frac{1}{\pi} \int_{-\infty}^{+\infty} d\om \; \text{Im} \; V^R(q,\om)
\; n(\om) \; G(k+q,i \ep_n+ \om+i \om_l)\; 
G(k+q,i \ep_n+ \om ) \; \Lm(k+q,i\ep_n+ \om, i\ep_n+\om +i \om_l)
\;\; . \;\; 
\eea

Taking into account that
\be
n(\om-i\ep_n) = f(\om) \;\;,\;\;
\ee
we also have

\bea 
I_1 =  \frac{1}{2 \pi i} \int_{-\infty}^{+\infty} d\om \; V(q,\om-i\ep_n-i\om_l)
\; f(\om- i \om_l) \; G(k+q,\om - i \om_l)\;  \nonumber \\
\left\{ G^R(k+q,\om) \; \Lm(k+q,\om-i\om_l, \om +i \de)
- G^A(k+q,\om ) \; \Lm(k+q,\om-i\om_l, \om -i \de) \right\}
\;\; . \;\;  \\
I_2 =  \frac{1}{2 \pi i} \int_{-\infty}^{+\infty} d\om \; V(q,\om-i\ep_n)
\; f(\om) \; G(k+q,\om + i\om_l )\;  \nonumber \\
\left\{ G^R(k+q,\om) \; \Lm(k+q,\om +i \de, \om+i\om_l)
- G^A(k+q,\om) \; \Lm(k+q,\om -i \de, \om+i\om_l) \right\}
\;\; . \;\;
\eea

We perform the analytic continuation
\be 
i\ep_n \rightarrow \ep -i\de \;\;, \;\; i\om_l  \rightarrow \om_0+i\de \;\;,
\;\;  i\ep_n + i\om_l  \rightarrow \ep + \om_0+i\de \;\;, \;\;
\ee
with both $\ep,\om_0$ {\em real}, which yields

\bea 
I_V = \frac{1}{\pi} \int_{-\infty}^{+\infty} d\om \; \text{Im} \; V^R(q,\om)
\; n(\om) \; G^R(k+q, \ep + \om)\; 
G^A(k+q,\ep+ \om ) \;  \Lm(k+q,\ep +\om-i\de, \ep+\om +i\de)
\;\; , \;\; 
\eea

\bea
2\pi i \; I_1 = \int_{-\infty}^{+\infty} d\om \; V(q,\om-\ep-\om_0) 
\; f(\om-\om_0) \; G(k+q,\om-\om_0-i\de) \;  \nonumber \\
\left\{ G^R(k+q,\om ) \; \Lm(k+q,\om-\om_0-i\de, \om +i \de)
- G^A(k+q,\om) \; \Lm(k+q,\om-\om_0-i\de, \om -i \de) \right\} 
 \;\; , \;\; 
\eea

\bea
2\pi i \; I_2 =  \int_{-\infty}^{+\infty} d\om \; V(q,\om-\ep)
\; f(\om) \; G(k+q,\om + \om_0+i\de )\;  \nonumber \\
\left\{ G^R(k+q,\om) \; \Lm(k+q,\om +i \de, \om+ \om_0+i\de )
- G^A(k+q,\om) \; \Lm(k+q,\om -i \de, \om+ \om_0+i\de ) \right\}
\;\; . \;\;
\eea

Combining $I_1$ and $I_2$ we have
\bea
2\pi i \; (I_1+I_2) = \int_{-\infty}^{+\infty} d\om \; V(q,\om-\ep) \; f(\om) 
\; K_0
\;\; ,\;\;
\eea
with
\bea
K_0 = G(k+q,\om-i\de) \;
\left\{ G^R(k+q,\om+\om_0 ) \; \Lm(k+q,\om-i\de, \om +\om_0+i \de)  \right.
\nonumber \\  \left.
- G^A(k+q,\om+\om_0) \; \Lm(k+q,\om-i\de, \om+\om_0 -i \de) \right\} 
\nonumber  \\  
+G(k+q,\om + \om_0+i\de )\;  
\left\{ G^R(k+q,\om) \; \Lm(k+q,\om +i \de, \om+ \om_0+i\de )  \right.
\nonumber \\  \left.
- G^A(k+q,\om) \; \Lm(k+q,\om -i \de, \om+ \om_0+i\de ) \right\}
\;\; . \;\;
\eea
We want $\Lm(k+q,\om -i \de, \om +i\de )$, which enters the formula 
for the conductivity. Taking $\om_0 \rightarrow 0$,
we see that the term $G^R G^A \; \Lm(k+q,\om -i \de, \om +i\de )$
is multiplied by a total {\em zero prefactor}, due to the opposite 
signs of the contributions from $I_1$ and $I_2$.
The only surviving contribution is
\bea
K_0 \rightarrow K_1 = G^R(k+q,\om+\om_0) \;  G^R(k+q,\om) \;
\Lm(k+q,\om +i \de, \om+ \om_0+i\de )  \nonumber \\
-G^A(k+q,\om+\om_0) \;  G^A(k+q,\om) \;
\Lm(k+q,\om -i \de, \om+ \om_0-i\de )
\;\; . \;\;
\eea
Now we use eq. (\ref{lapp}), and we recall that the derivative
$d f(\ep)/d \ep$ in eq. (\ref{condsi}) yields $\ep \simeq 0$ for low $T$
in $V(q,\om-\ep)$.
Then, using
$1/(x+iS)^2-1/(x-iS)^2=-4i \;x \; S/(x^2+S^2)^2$, with
$x=\om - \xi_{k+q}$ and $S$ from eq. ({\ref{sigm}), for the term $I_1+I_2$
we make the approximation 
\bea
\int_{-\infty}^{+\infty} d\om \; V(q,\om-\ep) \; f(\om) \;
\left[ \left(G^R(k+q,\om) \right)^2 - \left(G^A(k+q,\om) \right)^2\; \right]
\nonumber \\
\simeq  
- 4 \;i \; S \; \left(G^R(k+q,\om=0) G^A(k+q,\om=0) \right)^2 \; 
\int_{-C_0}^{+C_0} d\om \; V(q,\om) \; f(\om) \; (\om - \xi_{k+q})
 \;\; , \;\;  \label{apr12}
\eea
where the integration cutoff $C_0$ is of the order of $\ef$. Here we assumed
that the {\em main $\om$ dependence} comes from the integrand shown.
The product $(G^R \; G^A)^2$ acts as an additional cut-off for $|\om|>C_0$,
hence this energy range is omitted. 

To simplify the notation, we write
\be
\Lm(k,\ep) \equiv \Lm(k,\ep+i\de, \ep-i\de)  \;\;.\;\;
\ee
For the term $I_V$ we also make an approximation similar to the one
in eq. ({\ref{apr12})
\bea
\int_{-\infty}^{+\infty} d\om \; \text{Im} \; V^R(q,\om) \; n(\om) \;
G^R(k+q,\ep+\om) \; G^A(k+q,\ep+\om) \; \Lm(k+q,\ep+\om)  \nonumber \\
\simeq  
G^R(k+q,\ep)\;  G^A(k+q,\ep) \; \Lm(k+q,\ep) \;
\int_{-C_0}^{+C_0} d\om \; \text{Im} \; V^R(q,\om) \; n(\om) \;\; . \;\;
\eea

Now we introduce {\em approximate forms} for the functions $f(x)$ and $n(x)$.
Namely we consider
\bea
f(x) \rightarrow f_A(x)=x/(4T) \;, \;  
\text{for } |x|<2T \;, \; f_A(x)=\text{sgn}(x)/2 \text{ for }|x|\geq 2T \;\;,
\nonumber  \\   
n(x) \rightarrow n_A(x)=T/x \;, \;  
\text{for } |x|<2T \;, \; n_A(x)=\text{sgn}(x)/2 \text{ for }|x|\geq 2T \;\;.
\label{cfap}
\eea
The functions $f_A(x)$ and $n_A(x)$ are continuous and asymptotically exact
for $|x| \ll 2T$ and $|x| \gg 2T$. They differ from the original $f(x)$
and $n(x)$ mostly at $x=2T$. Namely $f_A(x=2T)=1/2 = (1/c_F) f(x=2T)$
and $n_A(x=2T)=1/2 = (1/c_B) n(x=2T)$, where $f(x=2T)=0.3808$ and
$n(x=2T)=0.6565$. The ``correction'' constants are
\be
c_F = 0.762   \;\;,\;\;  c_B = 1.31 \;\;. \label{cfb}
\ee

Using these $f_A(x)$ and $n_A(x)$ we obtain the analytical expressions
for $P(q)$ and $P_{12}(q)$ below. If we wish to consider the substitution
$f(x) \rightarrow f_A(x)$ and $n(x) \rightarrow n_A(x)$ at {\em face value},
we should take $c_F = c_B = 1$ {\em hereafter}. Else, we consider the values
given in eq. (\ref{cfb}), and we note that $c_F$ and $c_B$ are introduced
{\em by hand} in the following expressions, in order to compensate
for the discrepancy, due to the approximation in eqs. (\ref{cfap}),
around $x=2T$. Overall the difference between these
two cases has an upper limit of $c_B-1=0.31$ for the appropriate terms
in $P(q)$, $P_{12}(q)$ and $R_{1k}$ below.

Thus we obtain
\bea
I_V = G^R(k+q,\ep) \; G^A(k+q,\ep) \; \Lm(k+q,\ep) \; P(q)
\;\; , \;\;
\eea
\bea
P(q) = \frac{g}{\pi} \left\{ c_B \; \frac{2\;T\;h_q}{a_q} \;
\tan^{-1} \left(\frac{2\;T}{a_q \; h_q} \right) + 
\frac{h_q^2}{2} \;
\ln\left( \frac{(h_q \; a_q)^2+C_0^2}{(h_q \; a_q)^2+4 T^2} \right) \right\}
\;\; , \;\;
\eea
with $h_q=r+Dq^2$ and $a_q=a+\xi^2 q^2$.

When $\Lm(k,\ep)$ is inserted in eq. (\ref{condsi}) for $\si$,
the dominant momenta are $k \sim k_F$, with $k_F$ the Fermi 
momentum. In this way $v_k^2$ can be inserted in the integrand below,
and we obtain the following equation for $\Lm(k,\ep)$
\bea
\Lm(k,\ep) = 1 + \sum_q \left\{ n_i \; U_i^2(q) - P(q) \right\} \;
\Lm(k+q,\ep) \;  G^R(k+q,\ep) \;  G^A(k+q,\ep) \; 
\left( \frac{ {\bf v}_{k+q} \;{\bf v}_{k} }{ v_k^2 } \right)  \nonumber \\
+ \sum_q \left( G^R(k+q,0) \; G^A(k+q,0) \right)^2 \;
\left( \frac{ {\bf v}_{k+q} \;{\bf v}_{k} }{ v_k^2 } \right) \; P_{12}(q)
\;\; , \;\;
\eea
where
\be
P_{12}(q) = \frac{2 g \; S }{\pi} \;  \left \{ a_q \; h_q
\ln\left( \frac{(h_q \; a_q)^2+C_0^2}{(h_q \; a_q)^2+4 T^2} \right) 
+ c_F \; \left [ a_q \; h_q - \frac{ a_q^2 \; h_q^2}{2 T} \; 
\tan^{-1} \left( \frac{2 T}{a_q \; h_q} \right)  \right ]
\right \} \;\; . \;\;  \label{eqp12}
\ee
Further, we assume that, for $k \sim k_F$, $\Lm$ is very weakly dependent
on $|q| \ll |k|$, i.e. $\Lm(k+q,\ep) \simeq \Lm(k,\ep)$.
This assumption means that
$\Lm(k,\ep)$ is a {\em smooth} funtion of $k \sim k_F$, which is
{\em consistent} with what follows, and is common in related
derivations \cite{metz}. Also we note that, as far as the
integration over $q$ is concerned, the contribution from $G^R G^A$
is subleading compared to the other terms. As a consequence
\bea
\Lm(k,\ep) = \frac{1+Q_k}{1- R_k\; G^R(k,\ep) \;G^A(k,\ep)} \;\; , \;\;
\label{eqla}
\eea
where
\bea
R_k = \sum_q \left\{ n_i \; U_i^2(q) - P(q) \right\} \; 
\left( \frac{ {\bf v}_{k+q} \;{\bf v}_{k} }{ v_k^2 } \right) \;\; , \;\;
 \label{eqrk}
Q_k = \left\{ G^R(k,0) \; G^A(k,0) \right \}^2 \; 
\sum_q \left( \frac{ {\bf v}_{k+q} \;{\bf v}_{k} }{ v_k^2 } \right) \; P_{12}(q)
\;\; . \;\;
\eea


\vspace{0.6cm}

{\bf 4. Calculation of the conductivity}

\vspace{0.3cm}

Taking into account eqs. (\ref{condsi}),(\ref{eqla}), $\si$ is given by
\bea
\si= \frac{ e^2}{3\; \pi} \; \int_{-\infty}^{+\infty} \; d\ep \;
\frac{df(\ep)}{d\ep} \;
\sum_k \; v_k^2 \; \left\{ \frac{\left(1+Q_k\right) \; G^R(k,\ep) \; G^A(k,\ep)}
     {1- R_k \; G^R(k,\ep) \; G^A(k,\ep)} 
- \text{Re} \left( G^R(k,\ep) \right)^2  \right \}
\;\;.\;\;
\eea
This is the central result of this work. 
Considering the limit of low $T$ we have
\bea
\si= \frac{ e^2}{3\; \pi} \;
\sum_k \; v_k^2 \; \left\{
\frac{\left(1+Q_k \right) \;
  G^R(k,0) \; G^A(k,0) }{1- R_k \; G^R(k,0) \; G^A(k,0)} 
- \text{Re} \left( G^R(k,0) \right)^2  \right \}
\;\;.\;\;
\eea
Overall, this is a decent approximate formula, valid for
intermediate $T$ as well.
In the relevant terms $P(q)$ and $P_{12}(q)$ explicit $T^2$ terms were kept.
The derivative of the Fermi distribution was taken as a delta function, which
is also a reasonable approximation for intermediate $T$.

We write
\be
R_k=R_{1k}+R_{2k} \;\; ,\;\;
\ee
where
\be
R_{1k}=-\sum_q P(q)  \;\; , \;\;
\ee
\bea
R_{2k}=\sum_q  n_i \; U_i^2(q) 
\left( \frac{  {\bf v}_{k+q} \;{\bf v}_{k} }{ v_k^2 } \right)
+\sum_q P(q) 
\left( 1- \frac{ {\bf v}_{k+q} \;{\bf v}_{k} }{ v_k^2 }\right) \;\; . \;\;
\eea
Incidentally, we note
that the {\em transport} scattering rate, due to the impurities,
$\tau_{tr}^{-1}=\sum_q  n_i \; U_i^2(q) 
\left( 1 - {\bf v}_{k+q} \;{\bf v}_{k} / v_k^2  \right) $
comes from the term $R_{2k}$. 

Considering $|q| \ll |k|$, we have 
${\bf v}_{k+q} \; {\bf v}_{k} = v_k^2 + B_{1k} \; q + B_{2k} \; q^2 + ...$
(where $ B_{1k}, B_{2k}$ are coefficients of a Taylor expansion)
and the {\em dominant} contribution for the criticality parameter
$a \rightarrow 0$ comes from the term $R_{1k}$. This is the case because
higher powers of $q$ in the numerator of the integrand in eq. (\ref{eqrk}) 
yield terms {\em less} singular in the parameter $a$.

We evaluate $R_{1k}$. The interesting contribution, including negative
powers of $a$, arises from the low $T$ limit, with $2T < a_q \; h_q$. 
Hence we consider a minimum $q_T$ given by $2T = a_{q_T} \; h_{q_T}$.
As in \cite{gkfluc} we consider 
a maximum $q_{max}=1/2 \tau_o v_F$, where $v_F$ is the Fermi velocity.
Also we approximate the logarithm in $P_q$ as $l_0 \simeq \ln (C_0/a_0 h_0)$,
where $a_0=a_q,h_0=h_q$ with $q=q_{max}$.

Then in 3-D
\bea
R_{1k} = -\frac{g}{2 \pi^2} \left( c_B \; \frac{2 T^2}{\xi^3 \sqrt{a}}
\left \{ \tan^{-1} \left( \frac{\xi q_{max}}{\sqrt{a}} \right) 
-\frac{1}{\xi q_{max}} \right \}
+ q_{max}^3 \; l_0 \left \{ \frac{r^2}{3} + \frac{2 r D \; q_{max}^2}{5}
+\frac{D^2 \; q_{max}^4}{7}  \right \}  \right) \;\;,  \;\;
\eea
while in 2-D
\bea
R_{1k} = -\frac{g}{2 \pi} \left( c_B \; \frac{2 T^2}{\xi^2}
\left \{ \frac{1}{a} - \frac{1}{\xi^2 q_{max}^2} \right \}
+ q_{max}^2  \; l_0 \left \{ \frac{r^2}{2} + \frac{ r D \; q_{max}^2}{2}
+\frac{D^2 \; q_{max}^4}{6}  \right \}  \right) \;\; . \;\; \label{r1ga}
\eea

We note that, upon assuming the Gaussian regime 
$\xi^{2} \; a= \text{const.}$ \cite{revi,her,mil1}, 
there is {\em no} diverging factor in $R_{1k}$
for $a \rightarrow 0$. This possibility only arises if $\xi$ and $a$
are {\em independent} parameters - c.f. also \cite{gkfluc}.
We do not explicitly evaluate the integral in $Q_k$ of eq. (\ref{eqrk})
because it does not yield any diverging factor for $a \rightarrow 0$.
As discussed below, overall vertex corrections due to $V(q,\om)$ 
do not modify 
appreciably the conductivity in the vicinity of the critical point.

To further evaluate the conductivity, we assume a parabolic dispersion 
relation $\ep_k$ so that $v_k=k/m$, with $m$ the mass of the electrons,
as in eq. (18) in \cite{gkfluc}. Then, with $x=\ep_k-\ef$, $N_F$ the 
density of states at the Fermi level and now taking both
$R_k \rightarrow R_F$ and $Q_{12} \rightarrow Q_{F}$
{\em independent} of k and evaluated at $k=k_F$, we obtain
\bea
\si =  \frac{2 \; e^2 \; N_F}{3\; \pi \; m} \; \int_{-\ef}^{\infty} \; dx \;
(x+\ef) \left \{ \frac{1+Q_F}{x^2+S^2-R_F} + \frac{S^2-x^2}{(x^2+S^2)^2}
\right \}
\;\; . \;\;
\eea
This yields
\bea
\si =  \frac{2 \; e^2 \; N_F}{3\; \pi \; m} \;
\left\{ (1+Q_F)\frac{\ef}{S_0} \left[\frac{\pi}{2}
+\tan^{-1} \left( \frac{\ef}{S_0}\right) \right]
+\frac{(1+Q_F)}{2} \; \ln \left( \frac{E_0^2+S_0^2}{\ef^2+S_0^2} \right) 
+1 - \frac{1}{2} \ln \left( \frac{E_0^2+S^2}{\ef^2+S^2} \right)
\right\}
\;\; , \;\; \label{finsig}
\eea
with $E_0 \sim O(\ef)$ (the upper limit of integration was taken as
$E_0$ for the $\ln(...)$ terms, which are ultraviolet divergent) 
and $S_0=\sqrt{S^2-R_F}$. Of course, eq. (\ref{finsig}) is not exact, due to
the use of the parabolic dispersion instead of the actual crystalline one.
However, it is advantageous in that it allows to discern more clearly the
essential dependence on $a$ and $T$. Eq. (\ref{finsig}) can be simplified,
for reasons explained in the paragraph after next, with the result
\be
\si \simeq  \frac{2 \; e^2 \; N_F \; \ef}{3 \; m \; S_0}  \;\; .
\ee

These two expressions are very {\em similar} to eq. (18) in \cite{gkfluc}
(modulo a sheer numerical prefactor),
which includes a part of the vertex corrections due to impurity scattering,
as we explain in the following.
For reference, the final simplified expression for the conductivity 
in \cite{gkfluc},
given after eq. (18) therein, is 
$\sigma = 4 \pi \; e^2 N_F \ep_F / (m \; \sqrt{S^2 - u_o})$  (where
$u_o = n_{i} V_{i}^2$, with $V_{i}$ the typical value of the impurity
scattering potential $U_{i}(q)$).

Here, the vertex correction term 
$Q_F \propto S \;  [G^R(k_F,0) \; G^A(k_F,0)]^2 = 1/S^3 $, where $S$ in eq.
(\ref{sigm})
contains a {\em negative power law} of $a \rightarrow 0$ (times $T^2$). 
Hence $Q_F$ is {\em negligible}. The two remaining logarithmic terms in 
eq. (\ref{finsig}) practically cancel each other (the remainder is just
$x_1-x_2-O(x_1^2)+O(x_2^2)$, where $x_1=R_F /[\ef^2+S_0^2]$ and
$x_2=R_F /[E_0^2+S_0^2]$). Further,
the factor $R_F$, which also emanates from the vertex corrections, enters
in the combination $S^2-R_F$ in the final expression for the conductivity.
It does not modify in an essential manner the dependence on either $T$
or $a \rightarrow 0$.
Notably the {\em square} of $S$, yielding the main
$a$ and $T$ dependence, is combined with the linear in $R_F$ term.
Manifestly $R_F$ is less singular than $S^2$ for $a \rightarrow 0$
\cite{gkfluc}, and overall of smaller magnitude. In other words, as 
in \cite{gkfluc}, the main dependence of the conductivity $\si$ 
on $T$ and $a$ is due to the 
combination of the self-energy of eq. (\ref{sigm}) and of the vertex
corrections from impurity scattering. The contribution of the vertex 
corrections from the fluctuation potential $V(q,\om)$ is {\em not}
essential.

Here the resistivity is taken as $\rho=\rho_0 + A \; T^2$ and the specific
heat is $C= \gamma \; T$. 
We note that our theory yields a Kadowaki-Woods ratio
$A/\gamma^2$ which is constant for $a \rightarrow 0$ (possibly times
a ln($a$) term) in 3-D {\em only} \cite{gkfluc},
and this is consistent with experiments
\cite{bia, geg1, geg, naka}. 

\vspace{0.5cm} 

{\bf 5. Overview}

\vspace{0.3cm} 

We calculate the conductivity, including vertex corrections due to both 
critical ferromagnetic fluctuations and disorder, in a weakly
disordered metal close to a quantum critical point. We explicitly
show that no appreciable effect results due to the fluctuation part
of the vertex corrections. Our results are in very good
agreement with relevant experiments in several materials
\cite{pag,bia,gri,geg1,geg,but,shib,bali,naka,anal}, and
complement our previous calculation which did not explicitly consider
vertex corrections \cite{gkfluc} due to $V(q,\om)$. 
The characteristic Fermi liquid
$A \; T^2$ dependence for the resistivity, with a prefactor $A$ diverging as 
$a \rightarrow 0$, found therein thus remains valid.

\vspace{.6cm}
{\bf Appendix A : On the calculation of the scattering rate}
\vspace{0.3cm}

The derivation below follows that of \cite{gkfluc}, i.e. (I), 
and equation numbers refer
to (I) as well. In the limit $T \rightarrow 0$ the thermal function 
$X= \coth(\om/2T) \; + \; \tanh((\ep-\om)/2T)$ in eq. (I-4) becomes
$X=2$ for $2T < \om <\ep$, and $X=0$ for $\om<-2T$ and $\om>\ep$.
Then the integration over $\om$ - compare with eq. (I-7) - amounts to
\bea
2 \int_{2T}^{\ep} d\om \; \text{Im} \; V(q,\om) \; R(q,\om) \simeq  
g \ R_0 \; \ln\left( \frac{(h_q \; a_q)^2+\ep^2}{(h_q \; a_q)^2+ 4T^2} \right)
\simeq g \; R_0 \; \frac{\ep^2}{(h_q \; a_q)^2} \;\; , \;\;
\eea
for $ h_q \; a_q > \ep$. The rest of the algebra proceeds as in eq. (I-8) and
onwards. Thus the scattering rate scales like $\ep^2$ as well, 
as expected for the FL regime.

\vspace{.6cm}
{\bf Appendix B : The terms $R_V,R_1,R_2,L$ in eq. (\ref{eqint})}
\vspace{0.3cm}

The terms $R_1,R_2$ each contain a single propagator $G$. Hence, upon 
the final integration over momentum $k$ they both yield a small contribution.
This is the case because this integration is {\em similar} to an integration
over $\ep_k$ from $-\infty$ to $+\infty$, which can be taken as part of 
a contour 
integral closing at infinity. That contour can be taken such that the pole
of the $G$ in the integrand lies outside of it, and hence yields a zero
contribution. C.f. also ref. \cite{mahan}.

The term $R_V$ is due to the residue from the pole $z=z_0=-i \; a_q \; h_q$
of $V(q,z)$. Here both $G$'s enter the formula for the residue. However,
their poles are on the same semi-plane (i.e. in a combination $G^A G^A$),
and the argument for $R_1,R_2$ applies as well.

The term $L$ is the residue from the 2 poles $z=z_k,z_k^*$ of $\Lm(k,z)$
- c.f. eqs. (\ref{eqla}),(\ref{eqrk}) - with
\be
z_k = \xi_k + i \; W_k \;\;, \;\; W_k^2=S^2-R_k \;\; . \;\;
\ee
Considering the function 
\bea
H(z)=n(z) \; V(q,z)  \; G(k+q,i \ep_n+ z +i \om_l)\; 
G(k+q,i \ep_n+z ) \; R_{k+q} \;\;
\eea
we have 
\bea
L = \frac{H(z_{k+q})}{z_{k+q}-z_{k+q}^*} + \frac{H(z_{k+q}^*)}{z_{k+q}^*-z_{k+q}}
  \;\; . \;\;
\eea
This term is much smaller than $I_V$ because $|\text{Im} \; H(z)| 
\ll |\text{Re} \; H(z)| $.

\vspace{.3cm}
$^*$ e-mail : kast@iesl.forth.gr ; giwkast@gmail.com


\begin{thebibliography}{99}
\bibitem{fnw}
For the use of functions $f(\ep)$ and $n(\om)$ we follow ref. \cite{agd}.
\end{thebibliography}
\end{document}